\documentclass[twocolumn,floats,showpacs,amsmath,amssymb,prb]{revtex4}

\usepackage{graphicx}

\begin{document}

\title{Path-integral simulation of ice Ih: The effect of pressure}
\author{Carlos P. Herrero}
\author{Rafael Ram\'{\i}rez}
\affiliation{Instituto de Ciencia de Materiales de Madrid,
         Consejo Superior de Investigaciones Cient\'{\i}ficas (CSIC),
         Campus de Cantoblanco, 28049 Madrid, Spain }
\date{\today}

\begin{abstract}
The effect of pressure on structural and thermodynamic properties of ice
Ih has been studied by means of path-integral molecular dynamics
simulations at temperatures between 50 and 300 K. 
Interatomic interactions were modeled by using the effective q-TIP4P/F 
potential for flexible water.
Positive (compression) and negative (tension) pressures have been considered, 
which allowed us to approach the limits for the mechanical stability of this 
solid water phase.
We have studied the pressure dependence of the crystal volume, bulk
modulus, interatomic distances, atomic delocalization, and kinetic energy.
The spinodal point at both negative and positive pressures is derived from 
the vanishing of the bulk modulus. 
For $P < 0$, the spinodal pressure changes from --1.38 to --0.73 GPa in the 
range from 50 to 300~K.
At positive pressure the spinodal is associated to ice amorphization,
and at low temperatures is found between 1.1 and 1.3 GPa.
Quantum nuclear effects cause a reduction of the metastability 
region of ice Ih.
\end{abstract}

\pacs{31.70.Ks, 81.40.Vw, 65.40.De, 71.15.Pd} 


\maketitle

\section{Introduction}

Despite the large volume of experimental and theoretical work on
condensed phases of water, some of their properties still lack a complete
understanding. This is mainly due to the peculiar structure of liquid and 
solid water, where hydrogen bonds between contiguous molecules give rise to 
properties somewhat different than those of most known liquids and
solids (the so-called water ``anomalies'').\cite{ei69,pe99,fr00,ro96}

In the temperature region between 80 K and 250 K, ice Ih is the stable 
phase of water up to a pressure around 0.2 GPa (that increases for
increasing temperature). This range may seem rather narrow for some purposes, 
especially if one compares it with other types of materials different from
molecular solids.  However, this pressure region is large if
one takes into account the nature of the intermolecular interaction in
water, which gives rise to the presence of hydrogen bonds.
The region of mechanical stability of ice Ih, in which this phase is
metastable has been studied both experimentally and by computer
simulations. 
At 80 K, simulations\cite{sc95} have predicted it to be metastable down to a 
negative pressure (tension) of about --1.4 GPa, whereas for positive pressure 
(compression) its range of metastability is known to extend up to around
1 GPa, where ice Ih transforms into an amorphous phase (the
so-called high-density amorphous ice).\cite{mi84}
This pressure-induced amorphization of ice Ih is an issue that has
received in recent years a considerable amount of attention, along with
the existence of different amorphous phases which clearly differ in 
density.\cite{ts99,ts05,ne06}

Computer simulation of condensed phases of water at an atomic level 
has a long history, dating back to around 1970.\cite{ba69,ra71}
Since then, important efforts have been focused on the development 
and refinement of empirical potentials to describe both liquid and solid 
phases of water, so that nowadays a large variety of this kind of
potentials can be found in the literature.\cite{ma01,ko04,jo05,ab05,pa06,mc09}.
Many of them assume a rigid geometry for the 
water molecule, and some others include molecular flexibility either 
with harmonic or anharmonic OH stretches. 
In recent years, moreover, simulations of water using \textit{ab initio} 
density functional theory (DFT) have been carried out.\cite{ch03,fe06,mo08}
Nevertheless, it turns out that the hydrogen bonds in condensed phases 
of water are difficult to describe with presently available energy
functionals, which causes that some properties cannot be accurately reproduced 
by DFT calculations.\cite{yo09}  Some promising advances to improve the 
description of van der Waals interactions in water within the DFT formalism 
have been recently presented.\cite{wa11,ko11,ak11}

A limitation of {\em ab-initio} electronic-structure calculations is 
that they usually treat atomic nuclei as classical particles, not including 
quantum effects like zero-point motion.
These effects can be taken into account by using harmonic or quasiharmonic 
approximations for the nuclear motion, but  the precision of these
approaches is not easily estimated when large anharmonicities are present, 
as can be the case for light atoms like hydrogen.
To take into account the quantum character of atomic nuclei, the path-integral 
molecular dynamics (or Monte Carlo) approach has proved to be very useful,
since in this procedure the nuclear degrees of
freedom can be quantized in an efficient manner, thus including
both quantum and thermal fluctuations in many-body systems
at finite temperatures.\cite{gi88,ce95}
Thus, a powerful method can consist in combining DFT to determine the
electronic structure and path integrals to describe the quantum motion of
atomic nuclei.\cite{ch03,mo08} However, this procedure requires computer 
resources that would restrict enormously the number of state points that 
can be considered in actual calculations. 

The phase diagram of water is now known up to temperatures and pressures
on the order of 1000 K and hundreds of GPa, respectively. Present-day
interatomic potentials are able to predict rather accurately the range
of stability corresponding to the different known phases.\cite{sa04}
It is however important to estimate the influence of quantum nuclear
effects on the stability range of the different ice phases.
In this line, it is known that such quantum effects change the melting
temperature $T_m$ of ice Ih at ambient pressure by some degrees, as
manifested by the isotope effect on $T_m$.\cite{ra10}
This kind of quantum effects has been less studied for liquid and solid
water under external pressure.    In particular,
the spinodal line for ice Ih at negative pressures (which defines the limit
of mechanical stability of this water phase) has been calculated by
using classical molecular dynamics simulations.\cite{sc95}
Earlier studies of ice Ih using path-integral simulations have been 
carried out by using mainly effective potentials, and were
focused on structural and dynamic properties of the solid 
phase.\cite{ga96b,he05,he06b,pa08,mc09}

Quantum nuclear effects become more relevant for light atomic masses,
and are expected to be especially important in the case of hydrogen. 
Then, we pose the question of how quantum effects associated to the
lightest atom can influence the structural properties of a solid water
phase such as ice Ih, and in particular if these effects are relevant or
detectable for the solid at different densities, i.e. under different
external pressures. This refers to the crystal volume and interatomic
distances, but also to the mechanical stability of the solid.
In this context,
it is usually assumed that increasing quantum fluctuations monotonically
enhances the exploration of the energy landscape. However, in certain
regimes an increase in quantum fluctuations can lead to dynamical arrest,
as shown for glass formation.\cite{ma11}
A similar mechanism could also retard the transition from crystalline to
disordered (amorphous) phases.

In the present paper we study ice Ih by path-integral molecular 
dynamics (PIMD) simulations at different pressures and temperatures,
to assess the range of mechanical stability of this water phase, as well
as to analyze its structural properties. 
Interatomic interactions are described by the flexible q-TIP4P/F model, 
which has been recently developed and was employed to carry out PIMD 
simulations of liquid water\cite{ha09} and ice Ih at ambient 
pressure.\cite{ra10,he11}

 The paper is organized as follows. In Sec.\,II, we describe the
computational method and the model employed in our calculations. 
Our results are presented in Sec.\,III, dealing with the pressure
dependence of crystal volume, bulk modulus, interatomic distances, 
atomic delocalization, and kinetic energy of ice Ih.  Sec.\,IV includes 
a summary of the main results.

\section{Computational Method}

We employ here the PIMD method to obtain equilibrium properties of ice Ih
at different temperatures and pressures.
This method is based on an isomorphism between the actual quantum system
and a classical one, that appears after a discretization of the quantum 
density matrix along cyclic paths.\cite{fe72,kl90}
This isomorphism is in fact obtained by replacing each quantum particle by
a ring polymer consisting of $L$ (Trotter number) classical particles, 
connected by harmonic springs with temperature- and mass-dependent force
constant.
Details on this simulation technique can be found elsewhere.\cite{gi88,ce95}
We note that the dynamics employed in this method is fictitious and does
not correspond to the actual quantum dynamics of the real particles under
consideration, but it is used to effectively sample the 
many-body configuration space, giving precise results for the equilibrium 
properties of the quantum system.
An alternative way to derive equilibrium properties is the use of
Monte Carlo sampling, but this procedure has been found to require for the
present problem more computer resources than the PIMD method.
A particular advantage of the latter is that in this case the codes can be
more readily parallelized, an important factor for efficient use of modern
computer architectures.

Simulations of ice Ih have been carried out here in the 
isothermal-isobaric $NPT$ ensemble ($N$, number of particles; 
$P$, pressure; $T$, temperature), which allows us to obtain the equilibrium 
volume of the solid at given pressure and temperature.
We have employed effective algorithms for performing PIMD simulations 
in this statistical ensemble, as those described in the 
literature.\cite{ma96,tu98,tu02}
Sampling of the configuration space has been carried out at temperatures
between 50 K and 300 K, and pressures in the region of mechanical stability
of ice Ih (which at 50 K corresponds to the range from --1.4 to 1.1 GPa). 
Note that we have employed both negative (tension) and positive (compression) 
pressures in the simulations.
For comparison with results of PIMD simulations, some simulations of
ice Ih were also carried out in the classical limit, which is obtained
in our path integral procedure by setting the Trotter number $L$ = 1.

Our simulations were carried out on ice Ih supercells with periodic
boundary conditions.
To check the influence of finite-size effects on the results,
we considered orthorhombic supercells of two different sizes.
The smaller one included 96 water molecules and had parameters 
$(3a, 2 \sqrt{3} a, 2c)$, where $a$ and $c$ are the standard hexagonal 
lattice parameters of ice Ih, whereas
the larger supercell included 288 molecules and had parameters
$(4a, 3 \sqrt{3} a, 3c)$.
Results obtained for both types of supercells coincided within the error
bars due to the statistical uncertainty associated to the simulation
method.
The flexibility of the simulation cell was taken into account in the
$NPT$ simulations by treating the modules of the orthorhombic cell vectors 
as independent dynamic variables.
In the considered supercells, and before the PIMD simulations, 
proton-disordered ice structures were generated by a Monte Carlo procedure, 
in such a way that each oxygen atom had two chemically bonded and two H-bonded 
hydrogen atoms, and with a cell dipole moment close to zero.\cite{bu98}

The interatomic interactions have been modeled by the point charge,
flexible q-TIP4P/F model, recently developed to study liquid
water,\cite{ha09} and that was subsequently employed to study several
properties of ice,\cite{ra10,he11} as well as water clusters.\cite{go10}
In line with the arguments discussed by Habershon {\em et~al.},\cite{ha09}
there are some reasons to take this model potential for the present
study. First, most previous works that have considered quantum effects
in water phases have employed empirical potential models that were
parameterized on the basis of earlier classical simulations.
Then, quantum simulations using those models lead to ``double
counting'' of quantum effects.
Second, many of the empirical potentials previously used for quantum
simulations of condensed phases of water treat H$_2$O molecules as rigid 
bodies.\cite{he05,mi05,he06b} This can be convenient for computational
efficiency, but it disregards the important role of intramolecular
flexibility in the structure, dynamics, and thermodynamics of the
condensed water phases.\cite{ha09} 
Third, the significant anharmonicity of the O--H vibration of the water
molecule is taken into account by anharmonic stretches in the 
q-TIP4P/F potential, vs. the harmonic potentials employed in the
majority of simulations that considered quantum effects.

Technical details on the simulations presented here are the same as 
those employed and described in Refs.~\onlinecite{ra10,he11}.
In particular, the Trotter number $L$ was taken proportional to the 
inverse temperature ($L \propto 1/T$), so that $L T$ = 6000~K,
which allows one to keep roughly a constant precision in the PIMD results
at different temperatures.  Also,
the time step $\Delta t$ associated to the calculation of interatomic forces
was taken in the range between 0.1 and 0.3 fs, which was found to provide 
adequate convergence for the variables studied here.
For given temperature and pressure, a typical simulation run consisted of 
$5 \times 10^4$ PIMD steps for system equilibration, followed by 
$6 \times 10^5$ steps for the calculation of ensemble average properties.
At ambient pressure, the interatomic potential q-TIP4P/F predicts
melting of ice Ih at 251 K,\cite{ra10} but here we study this water 
phase up to 300 K, a temperature at which it was found to be metastable 
along our simulations.

From PIMD simulations one can obtain insight into the atomic delocalization
at finite temperatures. This includes a thermal (classical) delocalization,
as well as a delocalization associated to the quantum character of the
atomic nuclei, which is quantified by the extension of the quantum paths in
the path integral formalism.
For each quantum path of a given particle, one can define the
center-of-gravity (centroid) as
\begin{equation}
   \overline{\bf r} = \frac{1}{L} \sum_{i=1}^L {\bf r}_i  \, ,
\label{centr}
\end{equation}
where ${\bf r}_i$ is the position of bead $i$ in the associated ring
polymer.
Then, the mean-square displacement $\Delta_r^2$ of the atomic nuclei 
(H or O) along a PIMD simulation run is defined as
\begin{equation}
\Delta_r^2 =  \frac{1}{L} \left< \sum_{i=1}^L 
           ({\bf r}_i - \left< \overline{\bf r} \right>)^2
           \right>    \, ,
\label{delta2}
\end{equation}
where $\langle ... \rangle$ indicates a thermal average at a given 
temperature.
In connection with the quantum delocalization of a particle, in this
formalism it is interesting the spread of the paths associated to the
particle, which can be measured by the mean-square ``radius-of-gyration'' 
$Q_r^2$ of the ring polymers:
\begin{equation}
 Q_r^2 = \frac{1}{L} \left< \sum_{i=1}^L
             ({\bf r}_i - \overline{\bf r})^2 \right>    \, .
\label{qr2}
\end{equation}

\section{Results}

\subsection{Volume}

\begin{figure}
\vspace{-1.1cm}
\hspace{-0.5cm}
\includegraphics[width= 9cm]{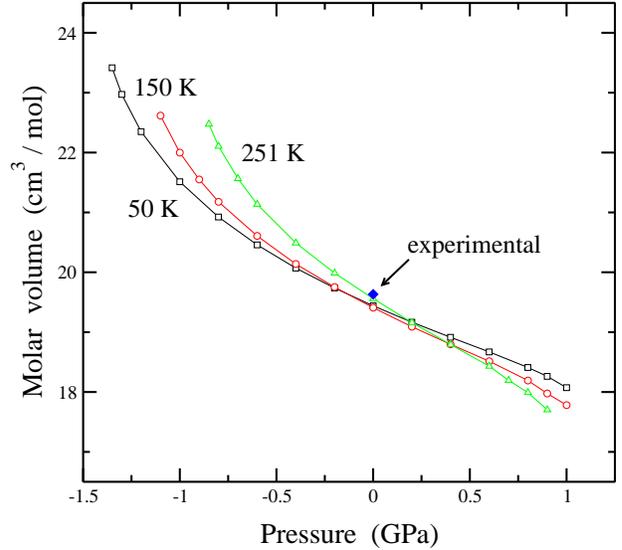}
\vspace{-1.0cm}
\caption{(Color online)
Molar volume of ice Ih as a function of pressure, as derived
from PIMD simulations at various temperatures: 50 K (squares),
150 K (circles), and 251 K (triangles).
Error bars are in the order of the symbol size.
Lines are guides to the eye.
A solid diamond represents experimental data from
Refs.~\onlinecite{gi47,da68} at atmospheric pressure and 273 K.
}
\label{f1} 
\end{figure}

We first present results for the equilibrium volume of ice Ih, as derived 
from our PIMD simulations at given pressure and temperature, in the
pressure region where the solid turned out to be mechanically stable.
These results are summarized in Fig.~1, where
we show the pressure dependence of the molar volume at three
different temperatures: 50, 100, and 251 K.
There are several important observations to make with regard to this figure.
At a given temperature, we observe the usual volume decrease for rising 
pressure, i.e., $d V/d P < 0$. 
In most of the pressure region considered, the slope $d V/d P$ becomes
less negative as the pressure is raised, which means $d^2 V/d P^2 > 0$. 
However, at about $P$ = 0.5 GPa we observe a change in the trend of the
first derivative as indicated by the presence of an inflection point 
with $d^2 V/d P^2 = 0$.
At higher pressures, the slope increases in absolute value until
reaching the limit of mechanical stability of the solid, close to 
$P$ = 1~GPa.
This change in the trend of the $P-V$ curve, with the appearance of an
inflection point seems to be related to the proximity of the amorphization
of the solid, with an important reduction in the molar 
volume.\cite{mi84,sc95,st04}
In fact, this amorphization is associated with a divergence in the
compressibility of ice Ih, where the derivative $d V /d P$ should diverge
to $-\infty$ (see below).

It is also remarkable that the volume-pressure curves cross at 
$P \sim$ 0.2--0.3~GPa.
At negative pressures we find that the larger volumes correspond to the
higher temperatures, as expected for the usual thermal expansion of solids.
However, at $P > 0$ one finds just the opposite trend: the higher volume
corresponds to the lower temperature, a fact associated to the negative 
thermal expansion occurring for ice Ih at low $T$. 
The potential TIP4P/F is able to reproduce this negative thermal expansion
at atmospheric pressure, as discussed elsewhere.\cite{he11}
At $P$ = 1 atm ($10^{-4}$ GPa), the line corresponding to 251 K in Fig.~1 
is still above those found for lower temperatures, but lies below them at 
higher (but relatively low) positive pressures.
This anomalous behavior of the crystal volume as a function of temperature
is due to the negative sign of the Gr\"uneisen parameter for TA vibrational 
modes.\cite{st04} 
For a mode in the $n$'th phonon branch with wave vector ${\bf q}$, this
parameter $\gamma_n({\bf q})$ is defined from the logarithmic
derivative of its frequency with respect to the crystal
volume:\cite{as76}
\begin{equation}
\gamma_n({\bf q}) = - \frac {\partial \ln \omega_n({\bf q})}
 {\partial \ln V} \, .
\end{equation}
Negative values of $\gamma_n({\bf q})$ for TA modes in ice have been 
related to the tetrahedral coordination of water molecules, and a negative
thermal expansion has been also observed in other solids with similar 
structures. In particular, it is well known for crystals with diamond 
and zinc blende structure.\cite{ev99}
In connection with this, negative values of $\gamma_n({\bf q})$ 
for TA modes were also found to be related with the pressure-induced 
amorphization of ice Ih.\cite{st04} In fact, they cause that the relations 
between elastic constants necessary for crystal stability are violated 
at a certain applied pressure, giving rise to mechanical instability of 
the solid.
In this line, instabilities in transverse acoustical modes
in ice Ih were also found at low temperatures from incoherent inelastic
neutron scattering.\cite{be99b}

For $P< 0$, the absolute value of the slope $d V/d P$ increases fast, and 
eventually diverges for a finite value of the pressure, which depends on
temperature. This is associated with a divergence of the compressibility of
the solid, which gives the limit for the mechanical stability of this ice
phase (spinodal point) at negative pressures.
On the other side, for positive pressures the stability limit at 
$P \sim$ 1~GPa is related to the amorphization of the solid. In fact, at
these pressures we find in the PIMD simulations a sudden reduction in the
volume of the solid, accompanied by the breaking and reordering of hydrogen
bonds (see below). 
Note that the possibility of studying a solid phase in metastable conditions,
close to a spinodal line is limited by the appearance of nucleation events,
which cause the breakdown of the crystal structure and the surge of the
equilibrium phase.
In this kind of atomistic simulations the probability of such nucleation 
events at low temperatures is relatively low, and the metastable range of the 
solid that can be explored is rather large. In fact, one can go to conditions 
close to the spinodal lines at positive and negative pressures, especially
at $P < 0$.

At 250 K ice Ih was found to be stable along our simulations up to
$P$ = 0.9~GPa. 
At such pressures, the H-bond network of ice Ih collapses and the crystal
transforms into the amorphous phase with an important volume decrease.
At $T$ = 251 K and $P$ = 1~GPa, this decrease amounts to about
18\% of the crystal volume.
Something similar happens at lower temperatures, and at 100 K we found a
volume reduction of 24\% in the amorphization process.
This is in line with the pressure-induced amorphization of ice Ih first
observed at $T$ = 77 K, and occurring at about $P$ = 1~GPa.\cite{mi84}
This water phase is the so-called high-density amorphous
ice.\cite{ts99,ga96}
Note that at higher $T$ (in the order of 250 K) amorphization
of ice Ih has not been experimentally observed, and a transition to 
ice III at $P \sim$ 0.2 GPa could be expected. Such transitions between
crystalline ice phases are usually not observed from the kind of atomistic 
simulations employed here, and other types of simulations employing 
thermodynamic integration would be required for this purpose.\cite{ra10} 
The important point here is that we obtain at each temperature the 
pressure range in which ice Ih is metastable with the employed potential
model, i.e., we are not discussing the transitions between equilibrium
(crystalline) phases, as they would appear in the phase diagram of
water. 

Our results are in line with those derived from classical molecular
dynamics simulations.\cite{sc95,ta02} In particular, Sciortino 
{\em et al.}\cite{sc95} employed the TIP4P potential, and found that 
the solid suffers reversible changes at
77 K for compression ($P > 0$) up to $\sim$ 1.5 GPa. At higher pressures, 
ice was found to become amorphous in an irreversible way, 
i.e. it remained in the amorphous phase when decreasing the pressure to
low values in the order of 1 atm.

The molar volume of ice Ih has been determined experimentally
in different ways, e.g., x-ray, optical, mechanical, calorimetric, 
acoustical, or nuclear methods (see Ref.~\onlinecite{fe06b} and
references therein). Measurements of different authors typically
deviate from each other by up to about 0.3\%.\cite{fe06b}
Our results at atmospheric pressure are in line with those of 
Refs.~\onlinecite{gi47,da68}, considered in Ref.~\onlinecite{fe06b} 
the most accurate determinations at normal pressure and $T$ = 273~K. 
These data are represented in Fig.~1 by a solid diamond.

\subsection{Bulk modulus and spinodal pressure}

The compressibility of liquid water and ice shows peculiar properties
associated to the hydrogen-bond network present in these condensed phases.
It turns out that their compressibility is smaller than what one
could expect at first sight from the large cavities present in their
structure, which could probably collapse under pressure without water
molecules approaching close enough to repel each other.
However, this is not the case, and for ice Ih in particular the
hydrogen bonds holding the crystal structure are known to be rather
stable, as revealed by the relatively high pressure necessary to break down
the H-bond network.\cite{mi84}

The isothermal compressibility $\kappa$ of ice, or its inverse the bulk
modulus [$B = 1/\kappa = - V ( {\partial P} / {\partial V} )_T$] can be
straightforwardly obtained from our PIMD simulations in the 
isothermal-isobaric ensemble. 
In this ensemble the isothermal bulk modulus
can be calculated from the mean-square fluctuations of the volume,
$\sigma_V^2 = \langle V^2 \rangle - \langle V \rangle^2$,
by employing the expression\cite{la80,he08}
\begin{equation}
       B = \frac{k_B T \langle V \rangle}{\sigma_V^2}   \; ,
\label{bulkm}
\end{equation}
$k_B$ being  Boltzmann's constant.
This expression was used earlier to obtain the bulk modulus
of different types of solids from path-integral
simulations.\cite{he00c,he08}

At atmospheric pressure we have found results for the bulk modulus of ice
Ih similar to those derived from experiments. In particular, at $T$ = 251 
and 300 K, we found $B$ = 9.1~GPa and 8.4~GPa, respectively.
There appears in the literature a dispersion of data for the isothermal
compressibility (or bulk modulus) of ice Ih at the melting temperature
and normal pressure.
The values of $B$ obtained here are in line with $B$ = 8.49~GPa at 
$T$ = 273 K,
corresponding to the equation of state derived by Feister and
Wagner\cite{fe06b} for this water phase from a set of experimental data.
It is difficult to estimate the precision of this value of the bulk modulus
due to the lack of error bars for the published data, and the
uncertainty may be large, as suggested by the dispersion in
experimental data obtained by different authors
(see Ref.~\onlinecite{fe06b} and references therein).
The uncertainty is even larger at lower temperatures, where we could
not find any direct experimental data for the isothermal compressibility.
The bulk modulus calculated from our PIMD simulations decreases for
rising temperature (a fact commonly found in solids), and this
behavior appears to be true for all studied pressures, except for
$P \gtrsim$ 1 GPa, where this general trend seems to be inverted
(see below).

\begin{figure}
\vspace{-1.1cm}
\hspace{-0.5cm}
\includegraphics[width= 9cm]{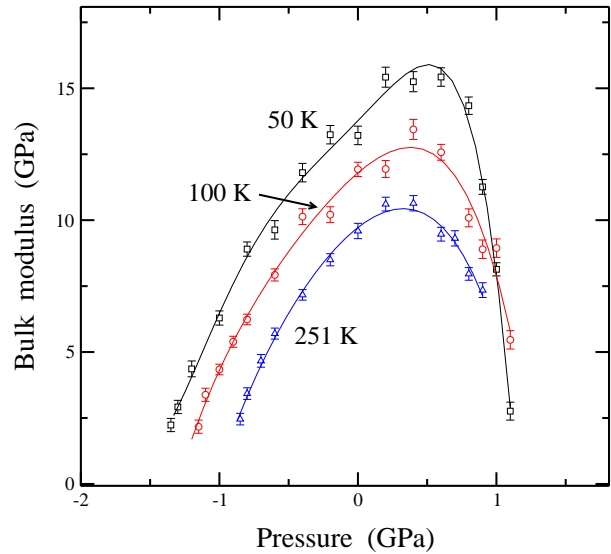}
\vspace{-1.0cm}
\caption{(Color online)
Pressure dependence of the bulk modulus of ice Ih as obtained from
quantum
PIMD simulations at several temperatures: 50 K (squares), 150 K
(circles),
and 251 K (triangles).
Error bars show the statistical uncertainty in the values of $B$
found from the simulations.
Lines are guides to the eye.
}
\label{f2}
\end{figure}

We will first discuss our results for negative pressure (solid under
tension).
For $P < 0$, the bulk modulus decreases as the pressure becomes more
negative (see Fig.~2), which means $\partial B / \partial P > 0$ at 
all temperatures
under consideration. Then, $B$ is expected to vanish at a
temperature-dependent pressure $P_s$, which corresponds to the spinodal
point defining the limit of mechanical stability of ice Ih at negative
pressures (where the compressibility diverges to infinity).

At a given temperature and near the spinodal pressure, our results for the
bulk modulus can be fitted to the expression
$B \sim (P - P_s)^{1/2}$, which allows us to obtain $P_s$
from a linear fit of $B^2$ vs $P$.  This pressure dependence of $B$ can be
understood in the following way.\cite{he03b}
At a temperature $T$, and calling $F$ the free energy, the spinodal point
is given by the condition
$\partial^2 F / \partial V^2 |_{V_s} = 0$,
which can be rewritten as
$\partial P / \partial V|_{V_s} = 0$.
Thus, close to $P_s$ one can write along an isotherm:
\begin{equation}
    P = P_s + C_2 (V - V_s)^2 + C_3 (V - V_s)^3 + ...
\label{pres1}
\end{equation}
with constants $C_2$ and $C_3$ independent of $V$. Assuming $C_2 \neq 0$
(which seems to be true here), one has
$B \approx 2 \sqrt{C_2} V_s (P - P_s)^{1/2}$  for small values of the
difference $P - P_s$
(for discussions on the case $C_2 = 0$, see
Ref.~\onlinecite{ma89,ma91,bo94}).
Such a pressure dependence for $B$, or its equivalent:
\begin{equation}
    P - P_s \sim (V - V_s)^2
\label{pres2}
\end{equation}
has been obtained earlier for ice and SiO$_2$ cristobalite, as well as
for rare-gas solids close to their stability limits.\cite{sc95,he03b}
This means in our case that the exponents describing the singularity are
consistent with those of the mean-field spinodal,\cite{sp82} given by
Eq.~(\ref{pres2}),   where $P_s$ and $V_s$ are
the values of the pressure and volume on the spinodal.
Thus, we have obtained $P_s$ at each temperature by fitting our results for
the bulk modulus at $P < 0$ to an expression $B \sim (P - P_s)^{1/2}$.
The results for $P_s$, as derived from our PIMD simulations, are shown in
Fig.~3 by solid circles.
$P_s$ becomes less negative as the temperature is raised, going from 
--1.38 GPa at 50 K to --0.73 GPa at 300 K.   

\begin{figure}
\vspace{-1.1cm}
\hspace{-0.5cm}
\includegraphics[width= 9cm]{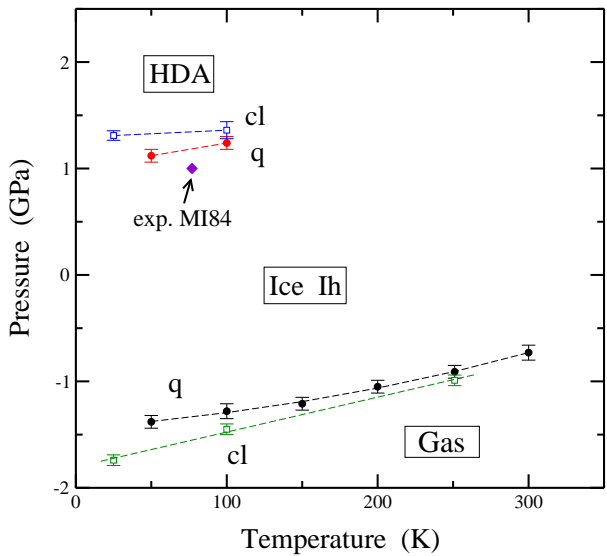}
\vspace{-1.0cm}
\caption{(Color online)
Calculated spinodal pressure for ice Ih as a function of temperature.
Solid circles are data obtained from PIMD simulations, whereas open
squares represent data yielded by classical simulations.
Labels `q' and `cl' indicate quantum and classical, respectively.
A solid diamond corresponds to the experimental point for amorphization
of ice Ih given in Ref.~\onlinecite{mi84} (denoted here as MI84).
Labels indicate the regions in the $P-T$ diagram where one finds the
different phases: ice Ih, high-density amorphous (HDA), and gas.
Lines are guides to the eye.
}
\label{f3}
\end{figure}

We now turn to the results at positive pressures (compression).
For $P >$ 0, the bulk modulus increases moderately until reaching 
a maximum value at about 0.3--0.5 GPa. This maximum
value of $B$ is lower for higher $T$, as shown in Fig.~2, and appears to be
slightly shifted to lower pressures as temperature increases.
Str\"assle {\em et al.}\cite{st05} measured the equation of state of ice Ih
at 145 K under hydrostatic pressure up to 0.5 GPa, and fitted their results
for the pressure dependence of the volume to a Murnaghan equation.
They obtained at zero pressure a bulk modulus $B$ = 9.85(47) GPa, 
somewhat lower than that derived from our simulations at 150 K: 
$B$ = 11.9~GPa.

At pressures larger than 0.5 GPa the bulk modulus derived from our PIMD 
simulations decreases rather fast, until
reaching at each temperature a certain pressure at which $B$ approaches
zero and ice Ih becomes mechanically unstable.
We will call $P_s'$ this spinodal pressure.
Close to $P_s'$ one has a relation between $B$ and $P$ 
similar to that given above, but valid here for $P_s' > P$, so that : 
$B \sim (P_s' - P)^{1/2}$.
Fitting our results to this expression we find that the bulk modulus 
extrapolates to zero for $P_s$ = 1.11 and 1.25 GPa, at 50 and 100 K, 
respectively. 
At higher temperatures, this extrapolation cannot be done from our PIMD
results in a reliable way, since the solid collapses into the amorphous
phase at pressures clearly lower than that corresponding to the metastability
limit of the crystal, where the bulk modulus is still far from vanishing.
This is due to a kinetic phenomenon such as the nucleation of a new phase,
which prevents close approach to the spinodal, as indicated above.
Thus, at 250 K ice Ih was metastable in our PIMD simulations only
up to $P$ = 0.9 GPa, where the bulk modulus is still larger than 7 GPa,
but the ice crystal transforms into the amorphous phase with an
appreciable volume decrease.
Arguments equivalent to those presented here for the ice instability 
associated to a vanishing bulk modulus, have been given in connection with 
the elastic moduli of this hexagonal crystal.\cite{ts92}

In Fig.~3 we present values of the spinodal pressure $P_s'$, as 
derived from our PIMD simulations, along with those corresponding to 
negative pressures, $P_s$ (represented as solid circles). 
In both cases the spinodal pressure increases as the temperature is raised.
At positive pressure, we show only results up to 100 K, as at higher $T$
the extrapolation employed to determine $P_s'$ is unreliable, as explained
above. These results are compared with those derived from classical
molecular dynamics simulations (open squares).
At positive pressure, $P_s'$ derived from classical simulations is
larger than that found from PIMD, and the opposite occurs for the
spinodal $P_s$ at negative pressures. In both cases, quantum and
classical results are found to lie closer as temperature is raised.
Altogether we find that quantum effects reduce the metastability range
of ice Ih. This means that at low temperature the pressure region in
which this water phase is metastable is reduced by about 0.3 GPa, 
both at negative and positive pressures.
For comparison with the calculated spinodal, we also show in Fig.~3 
the value of $P$ = 1 GPa given by Mishima {\em et al.}\cite{mi84}
for the laboratory amorphization pressure at 77 K (solid diamond).
Spinodal pressures for ice Ih have been calculated earlier by Sciortino
{\em et al.}\cite{sc95} from classical molecular dynamics simulations with
the TIP4P potential. These authors found 
at low temperatures values close to the results of our classical
simulations, but at high $T$ they obtained values somewhat higher (less
negative) than those found here, e.g., at 250 K they found 
$P_s = -0.75$~GPa vs our value of --0.99 GPa. This difference is 
presumably due to differences between the interatomic potentials 
employed in both calculations, as the TIP4P deals with rigid water 
molecules and the q-TIP4P/F takes into account their flexibility, 
allowing for bending and stretching of the intramolecular bonds.
 Simulations of the amorphization of ice Ih using classical 
molecular dynamics were also carried out by Tse and Klein,\cite{ts87,ts90} 
who found ice amorphization at pressures around 1.2--1.3 GPa at
temperatures between 80 and 100 K, close to the spinodal $P_s'$ obtained 
here. 

We remark that values of the bulk modulus calculated from PIMD simulations 
in the isothermal-isobaric ensemble show relative error bars larger than 
those obtained for other variables (e.g., molar volume, kinetic energy, or
interatomic distances), as a consequence of the statistical uncertainty 
in the volume fluctuations $\sigma_V$, employed to calculate the bulk modulus.
An alternative way to derive $B$ consists in calculating numerically the
derivative $\partial V / \partial P$ from the $P$--$V$ equation of state at
temperature $T$. We have checked that this method gives results that agree
within error bars with those derived from the volume fluctuations through
Eq.~(\ref{bulkm}). However, the calculation based on the pressure 
derivative of the volume may be less reliable in the pressure region where 
$B$ changes fast, as happens close to $P$ = 1~GPa (see Fig.~2).

\subsection{Interatomic distances}

\begin{figure}
\vspace{-1.1cm}
\hspace{-0.5cm}
\includegraphics[width= 9cm]{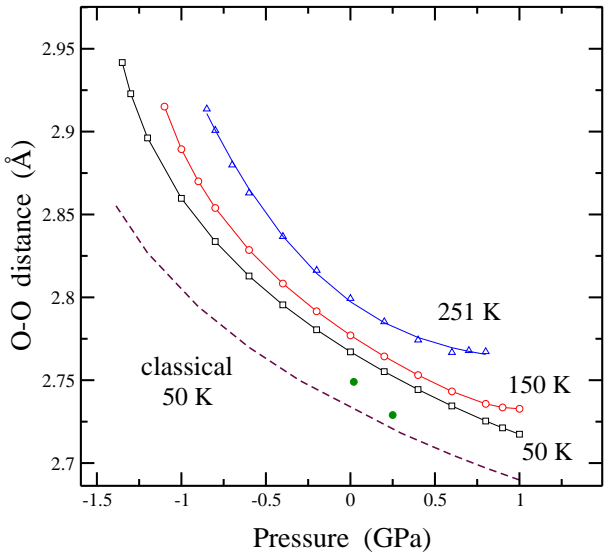}
\vspace{-1.0cm}
\caption{(Color online)
Mean distance between oxygen atoms in nearest-neighbor water
molecules in ice Ih, as a function of pressure.
Symbols represent results of PIMD simulations at different
temperatures:
50 K (squares), 150 K (circles), and 251 K (triangles).
Lines are guides to the eye.
The dashed line represents the results of classical simulations
at 50 K.
Solid circles represent O--O distances derived from the structural data
obtained by Str\"assle {\em et al.}\cite{st05} from neutron diffraction
experiments of D$_2$O ice Ih at 145 K.
}
\label{f4}
\end{figure}

In this section we show results for interatomic distances
in ice Ih between atoms in the same and adjacent molecules.
This can shed light on the structural changes suffered by the 
crystal when temperature and/or pressure are modified.
We first show in Fig.~4 the mean distance between oxygen atoms in
neighboring molecules as a function of external pressure.
Open symbols represent results of PIMD simulations at
different temperatures. In each case, we give values of $d$(O--O) in the
region where the crystal was found to be metastable in the PIMD
simulations. This includes, as in previous figures, positive and negative
values of $P$.
For increasing pressure, $d$(O--O) is reduced, as could be expected from the
associated volume decrease ($d V / d P < 0$).  
Nevertheless, contrary to the case of the crystal volume, the isothermal lines
corresponding to the O--O distance do not cross each other, so that
at higher temperature one finds a larger distance in the whole metastability
region of ice Ih. 

 For an homogeneous and isotropic expansion (or contraction) of the
crystal, one expects for small volume changes: 
$\Delta V / V = 3 \Delta l / l$, $l$ being any distance in the solid,
and one would expect this relation to be fulfilled, in particular,
for the intermolecular distances in the solid.
However, although this relation is roughly followed in the parameter range
considered here, it is obvious that it cannot be strictly accomplished,
given the apparent different trends of the curves shown for crystal volume
and O--O distance in Figs.~1 and 4.
In particular, volume curves at different temperatures cross each other,
whereas those corresponding to the O--O distance follow a regular pattern
without crossings.  
The molar volume at low temperature ($T$ = 50 K) decreases by 7.0\% when
going from atmospheric pressure to $P$ = 1 GPa, from where one would expect
a relative change in the average O--O distance of 2.3\%. This value is
somewhat higher than the change actually obtained from the simulations,
i.e., 1.8\%.
On the other side, at negative pressure we find a relative volume increase
of 20.4\% at $P$ = -1.35 GPa (close to the spinodal pressure $P_s$), which 
would correspond to a relative change in the O--O distance of 6.8\%. In fact, 
we find from the simulations an increase in the O--O distance of 6.3\%,
slightly smaller than that expected for an homogeneous and isotropic
crystal expansion.
This indicates that expanding or contracting the ice crystal is
accompanied by a distortion of the tetrahedral network of water molecules.
This kind of distortion is also related to the negative thermal expansion
of ice Ih occurring at low temperatures, and that was earlier found to be 
reproduced by PIMD simulations with the q-TIP4P/F potential.\cite{he11} 

For comparison with the results of PIMD simulations, we also show in
Fig.~4 results of classical molecular dynamics simulations with the
same interatomic potential, which are shown for $T$ = 50 K by a 
dashed line.
At atmospheric pressure the average O--O distance obtained in the
quantum simulations is 0.032 \AA\ larger than that derived from
classical simulations, which means a difference of 1.2\%.
This difference increases at negative pressures, and at $P = -1.2$ GPa
and $T$ = 50 K we find an increase of 0.069 \AA\ due to quantum
effects, i.e., a 2.4\% of the O--O distance.
These numbers are consistent with the volume increase associated to 
quantum effects. In fact, at $P$ = 1 atm the quantum result for the 
molar volume is 3.4\% larger than the classical one, vs an increase 
of 7.2\%  obtained at $P = -1.2$ GPa (see Ref.~\onlinecite{he11} for
results at ambient pressure and several temperatures).
In Fig.~4 we also display O--O distances derived from the structural
data obtained by Str\"assle {\em et al.}\cite{st05} from neutron diffraction
experiments of D$_2$O ice Ih at 145 K (solid circles).
Note that O--O distances in ice may change with the hydrogen 
isotope,\cite{he11} so that these data for D$_2$O ice at 0.02 and 
0.25~GPa could differ from those corresponding to H$_2$O ice, but in
any case they are useful for comparison with our PIMD results.

\begin{figure}
\vspace{-1.1cm}
\hspace{-0.5cm}
\includegraphics[width= 9cm]{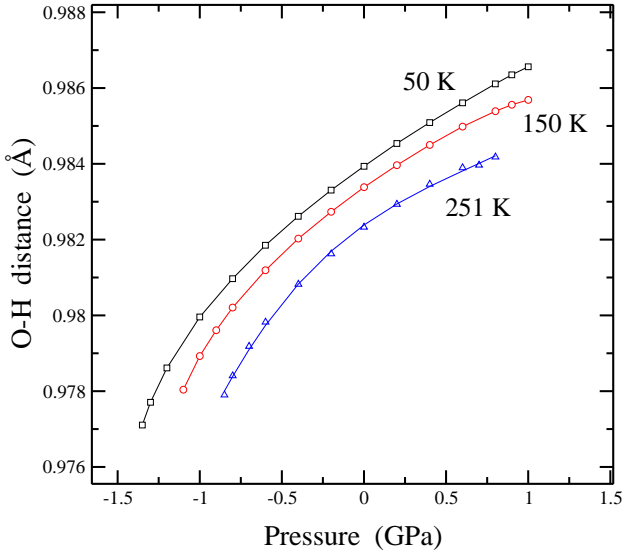}
\vspace{-1.0cm}
\caption{(Color online)
Pressure dependence of the mean intramolecular O--H distance,
as derived from simulations of ice Ih. Open symbols represent
results of PIMD simulations at different temperatures:
50 K (squares), 150 K (circles), and 251 K (triangles).
Error bars are in the order of the symbol size.
Lines are guides to the eye.
}
\label{f5}
\end{figure}

We now consider the interatomic distances inside water molecules
in the ice crystal, as derived from our simulations at different
pressures.
In Fig.~5 we have plotted the mean intramolecular O--H distance
as a function of pressure at three temperatures.
Concerning this figure, there appear two results that should be emphasized. 
First, at a given pressure we observe that the O--H distance decreases as
temperature is raised, contrary to the usual expansion of atomic bonds for
increasing $T$.
Second, at a given temperature, we find that the O--H bond distance
increases for rising pressure, which may seem also surprising given the
pressure-induced volume contraction.
Both facts are due to the characteristic structure of ice with hydrogen bonds
connecting contiguous water molecules, and can be explained by the same 
argument: An increase in the
intramolecular O--H distance is associated to a weakening of the O--H bond,
and is caused by a hardening of the intermolecular H bond.
For rising $T$, molecular motion is enhanced, so that
hydrogen bonds become softer, and the average intermolecular O--H
distance increases (in line with the rise in O--O distance shown above 
and in Ref.~\onlinecite{he11}). This gives rise to a strengthening of 
the intramolecular
O--H bonds, with the associated decrease in interatomic distance in
water molecules.  This hardening of intramolecular O--H bonds for rising 
temperature has been observed
experimentally and reported in the literature.\cite{ny06}
For increasing pressure at a given temperature, we find something similar:
the intermolecular O--O distance is reduced as the pressure rises
(see Fig.~4), in such a way that the intermolecular H bonds become
stronger, causing weaker and longer intramolecular O--H bonds. 
The rise in O--H distance derived from our simulations in the range
between 0 and 1 GPa is consistent with an increase of about
$2 \times 10^{-3}$ \AA/GPa, given in Refs. \onlinecite{ho72,kl84} for
high-pressure ice phases, although smaller values have been found 
in Ref. \onlinecite{be94}.

Note, however, that the change of intermolecular O--O distance with
pressure is much larger than that of the intramolecular O--H distance.
Thus, at 50 K the change in $d$(O--O) in the whole metastability region of
ice Ih (from $P$ = --1.38 to 1.12 GPa) is about 8\%, whereas the change in
intramolecular O--H distance is about 1\%.
This is indeed a consequence of the direct effect of pressure in
changing the intermolecular distance, to be compared with an indirect 
effect on the intramolecular O--H bond strength caused by intermolecular 
H bonds. 

We finally note that the intramolecular O--H
distance increases appreciably due to quantum nuclear effects. In fact,
a classical simulation at $T$ = 50 K and ambient pressure yields a mean
O--H distance of 0.969 \AA, to be compared with 0.984 \AA\ derived from
PIMD simulations, which means an increase of 1.5\% in the bond length.
This difference is much larger than the temperature-induced change in
$d$(O--H) shown in Fig.~5.
However, the increase due to quantum motion is rather constant in the
whole pressure range studied here, i.e. at negative and positive
pressures we find a rise in O--H distance of about 1.5\%.

\subsection{Atomic delocalization}

\begin{figure}
\vspace{-1.1cm}
\hspace{-0.5cm}
\includegraphics[width= 9cm]{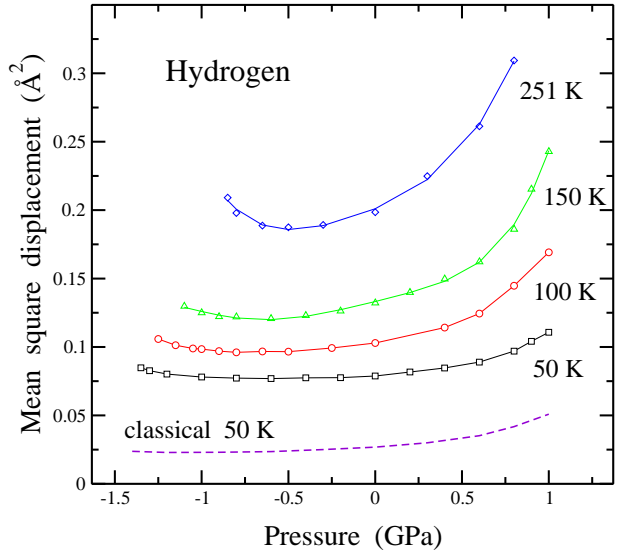}
\vspace{-1.0cm}
\caption{(Color online)
Pressure dependence of the spatial delocalization of hydrogen nuclei
(protons) in  ice Ih, as derived from PIMD simulations.
Symbols indicate the mean-square displacement
$\Delta_r^2$ at several temperatures:
50 K (squares), 100 K (circles), 150 K (triangles), and 251 K (diamonds).
Lines are guides to the eye.
The dashed line is the mean-square displacement of hydrogen derived from
classical molecular dynamics simulations at 50 K.
}
\label{f6}
\end{figure}

Here we present results for the atomic mean-square displacements  
$\Delta_r^2$ defined in Eq.~(\ref{delta2}). 
One expects that this spatial delocalization will be larger for hydrogen 
than for oxygen, due to the smaller mass of the former.
In Fig.~6 we display values of $\Delta_r^2$ for hydrogen in ice Ih as
a function of pressure. Data were derived from PIMD simulations at
different temperatures.
At a given pressure, $\Delta_r^2$ increases as the temperature is raised,
in the whole region where the solid is found to be metastable.
At 50 K (the lowest temperature shown here), $\Delta_r^2$ changes very
little as a function of pressure, except close to $P$ = 1 GPa, where it
increases by a small amount as one approaches the spinodal point 
(which at this temperature was found to be $P_s'$ = 1.12 GPa; see above).
At higher temperatures, changes in $\Delta_r^2$ as a function of pressure
are more appreciable. 
At all temperatures considered, we find an increase in the atomic
delocalization when approaching one of the spinodal points, at positive
or negative pressure. The increase at positive pressure (close to
amorphization of the solid) is clearly larger than that found at negative
pressures.
In Fig.~6 we also present the mean-square displacement of hydrogen
atoms in Ih, as derived from classical simulations at 50 K (dashed line). 
These results are clearly lower than those of the PIMD
simulations at the same temperature. Thus, at ambient pressure the
classical result is $\Delta_r^2 = 2.68 \times 10^{-2}$ \AA$^2$, to be
compared with $\Delta_r^2 = 7.88 \times 10^{-2}$ \AA$^2$ found in the
quantum simulations.

\begin{figure}
\vspace{-1.1cm}
\hspace{-0.5cm}
\includegraphics[width= 9cm]{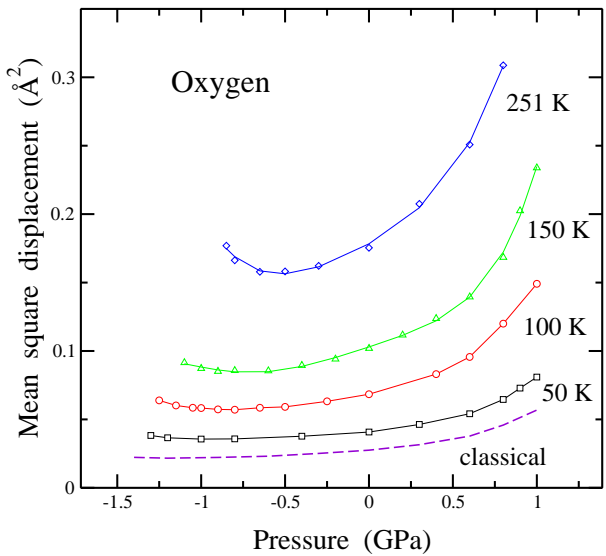}
\vspace{-1.0cm}
\caption{(Color online)
Pressure dependence of the spatial delocalization of oxygen nuclei,
as derived from PIMD simulations.
Symbols represent the mean-square displacement $\Delta_r^2$ at several
temperatures:
50 K (squares), 100 K (circles), 150 K (triangles), and 251 K (diamonds).
Lines are guides to the eye.
The dashed line is the mean-square displacement of oxygen derived from
classical molecular dynamics simulations at 50 K.
}
\label{f7}
\end{figure}

For oxygen we find mean-square displacements that change with pressure in
a way similar to those of hydrogen. This is shown in Fig.~7 at the same
temperatures as in Fig.~6 for hydrogen.
At atmospheric pressure, we find for oxygen 
$\Delta_r^2 = 4.08 \times 10^{-2}$ \AA$^2$ at 50 K vs 0.175 \AA$^2$  
at 251 K. Values of $\Delta_r^2$ are larger for hydrogen than for oxygen, 
but the difference between both decreases for increasing temperature.
Thus, at $T$ = 50 K and $P$ = 1 atm, $\Delta_r^2$ for hydrogen is 
1.93 times the mean-square displacement for oxygen, whereas this 
ratio decreases to 1.13 at 251 K.
This is in line with a larger quantum delocalization for hydrogen, caused by
its lower mass, and the effect of the mass becomes less relevant as
temperature increases.

Note that the mean-square displacement $\Delta_r^2$ of a given atomic
nucleus can be decomposed into a contribution $Q_r^2$ coming from the
spread of the quantum paths [see Eq.~(\ref{qr2})], and another one, 
$C_r^2$, given by the spatial displacement of the centroid 
$\overline{\bf r}$ of the paths.\cite{gi88,he11}
At 50 K and 1 atm, $Q_r^2$ represents a 62\% of $\Delta_r^2$ for hydrogen 
and a 25\% in the case of oxygen, which reflects the fact that the quantum
contribution to the atomic delocalization is more important in the case of 
hydrogen.
This can also be seen by directly comparing values of $Q_r^2$ for the
different atomic species, which result to be 4.8 times larger for H
than for O at 50 K and atmospheric pressure.  

At low temperatures, the quantum
contribution (spreading of the paths), $Q_r^2$, dominates the spatial
delocalization of hydrogen, since the centroid displacement, $C_r^2$,
converges to zero as $T \to 0$ K.
The opposite happens in the high-temperature limit (unreachable here for 
stability reasons), where the quantum contribution $Q_r^2$ eventually 
disappears, as corresponds to the classical limit.
 We note in passing that at low temperatures the quantum paths associated to
hydrogen have an average extension of about 0.15 \AA,
much smaller than the H--H distance in a water molecule, thus justifying 
the neglect of quantum exchange between protons in the PIMD simulations.

In Fig.~7 we also show mean-square displacements of oxygen atoms, as
derived from classical simulations at 50 K (dashed line). As one could
expect, these values are smaller than those derived from PIMD
simulations, but in the case of oxygen the difference between both sets
of results, although nonnegligible, is not so important as for hydrogen. 
In fact, at ambient pressure we found for oxygen
$\Delta_r^2 = 2.75 \times 10^{-2}$ \AA$^2$ in classical simulations
vs $4.07 \times 10^{-2}$ \AA$^2$ derived from PIMD simulations.
Note that the classical result is similar to that found for hydrogen at
the same pressure and temperature, but the quantum $\Delta_r^2$ for H
is clearly larger than that of O atoms.
Also, quantum effects on the mean-square displacement for oxygen
decrease rapidly as temperature is raised. Thus, at $T$ = 150 K and
$P$ = 1 atm we find in the classical simulations
$\Delta_r^2 = 0.091$ \AA$^2$, to be compared with
$\Delta_r^2 = 0.102$ \AA$^2$ derived from PIMD simulations, i.e.,
quantum effects increase $\Delta_r^2$ by a 12\% at this 
temperature.

\subsection{Kinetic energy}

The kinetic energy of atomic nuclei in a solid or molecule
depends on the mass and delocalization of the considered nucleus. 
Thus, although in a classical approach, each degree of freedom contributes
to the kinetic energy by an amount that depends only on temperature 
($k_B T / 2$, as given by the equipartition principle), 
in a quantum approach the kinetic energy, $E_k$, gives information on 
the environment and interatomic interactions seen by the considered
particle.  In particular, 
a typical quantum effect related to the atomic motion in solids is
that the kinetic energy at low temperature converges to a finite value
associated to zero-point motion, contrary to the classical result where
$E_k$ vanishes at 0 K.
Path integral simulations allow us to calculate the kinetic energy
of the quantum particles under consideration, which is related to 
the spread of the quantum paths. In fact, for a particle with a certain
mass and at a given temperature, the larger the mean-square 
radius-of-gyration of the paths, $Q_r^2$, the smaller the kinetic energy,
in line with the expectancy that a larger quantum delocalization causes a
reduction in the kinetic energy.\cite{gi88,he11}
Here we have calculated $E_k$ by using the so-called virial estimator, 
which has an associated statistical uncertainty appreciably lower
than the potential energy of the system.\cite{he82,tu98}

\begin{figure}
\vspace{-1.1cm}
\hspace{-0.5cm}
\includegraphics[width= 9cm]{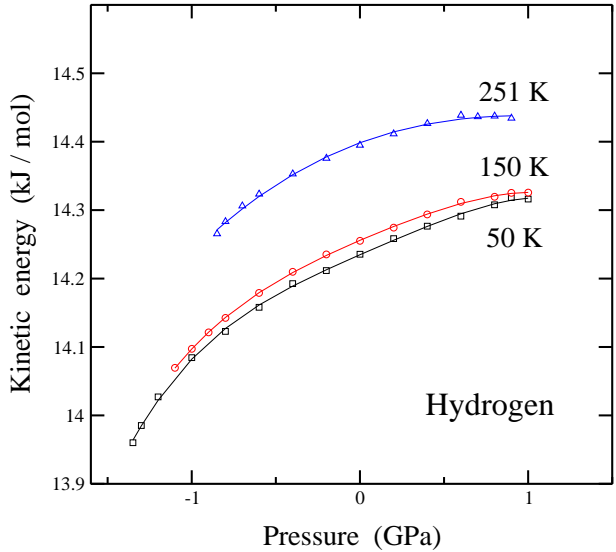}
\vspace{-1.0cm}
\caption{(Color online)
Kinetic energy of hydrogen in ice Ih as a function of pressure.
Open symbols indicate results derived from PIMD simulations
at different temperatures: 50 K (squares), 150 K (circles), and
251 K (triangles).
Error bars are less than the symbol size.
Lines are guides to the eye.
}
\label{f8}
\end{figure}

We present separately the kinetic energy of oxygen and hydrogen atoms in 
ice Ih.  In Fig.~8 we display $E_k$ for hydrogen as a function 
of pressure at three different temperatures, as derived from our PIMD
simulations. At each considered temperature, $E_k$ increases as pressure
rises, corresponding to an overall increase of vibrational frequencies.
$E_k$ also increases with temperature, but the data for $T =$ 50 and 100 K
(not shown in the figure)
are almost indistinguishable, since at these temperatures the hydrogen
vibrations are nearly in their ground state. Higher vibrational modes are
excited at higher temperatures, and at 251 K we observe an increase in
$E_k$ larger than 0.1 kJ/mol in the whole metastability range of ice Ih.
Values of $E_k$ calculated at $P$ = 1 atm coincide with those presented
earlier from PIMD simulations in the range from 210 to 290 K.\cite{ra11}

The kinetic energy of hydrogen atoms in ice Ih was obtained by Reiter
{\em et al.}\cite{re04} at ambient pressure from inelastic neutron 
scattering experiments.
These authors found at 269 K a kinetic energy $E_k$ = 12.7 kJ/mol,
somewhat lower than those found here at $P$ = 1~atm in the temperature 
range from 250 to 290 K, i.e., $E_k$ for hydrogen between 14.39 and 
14.44~kJ/mol.

\begin{figure}
\vspace{-1.1cm}
\hspace{-0.5cm}
\includegraphics[width= 9cm]{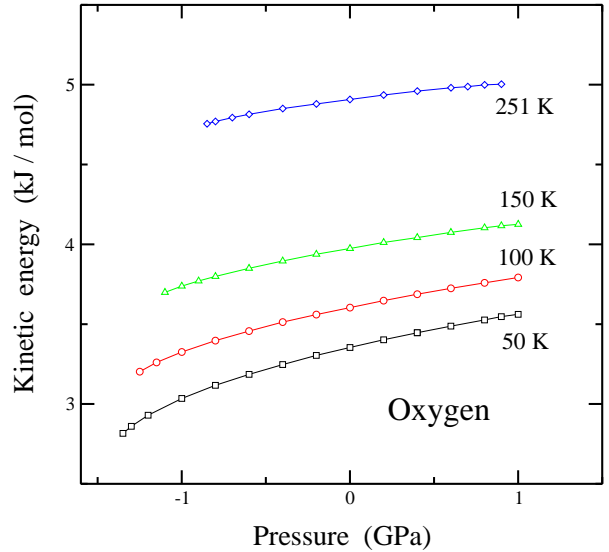}
\vspace{-1.0cm}
\caption{(Color online)
Kinetic energy of oxygen atoms in ice Ih as a function of pressure.
Open symbols indicate results derived from PIMD simulations
at different temperatures: 50 K (squares), 100 K (circles),
150 K (triangles), and 251 K (diamonds).
Error bars are less than the symbol size.
Lines are guides to the eye.
}
\label{f9}
\end{figure}

In Fig.~9 we show the pressure dependence of the kinetic energy of
oxygen in ice Ih at four temperatures.
As for hydrogen, for oxygen $E_k$ also grows for increasing pressure at 
a given temperature.  However, now the isothermal lines at 50 and 100 K are
clearly distinct by more than 0.2 kJ/mol, as a consequence of the higher 
mass of oxygen, which produces a lower vibrational energy and an 
appreciable excitation of vibrational modes at 100 K.
From the results for the kinetic energy of H and O shown in Figs.~8 and
9, we observe that at atmospheric pressure and low temperatures, $E_k$ for 
a hydrogen atom in ice Ih is about 4 times larger than that of an oxygen 
atom.  At 50 K the kinetic energy of hydrogen increases from 13.96 kJ/mol at 
$P$ = --1.35 GPa to 14.32 kJ/mol for 1~GPa. This represents an increase 
of 0.36 kJ/mol, i.e., 2.6\% in the whole pressure range. 
For oxygen, the increase in $E_k$ in the same region is about twice
that of hydrogen ($\Delta E_k$ = 0.75 kJ/mol), but its relative change 
is much more important for O (a 26.5\%).

Close to the amorphization pressure of ice we observe a slow increase in 
the kinetic energy for both hydrogen and oxygen, without any appreciable 
change in its trend. This
contrasts with the trend observed for the atomic mean-square displacements
$\Delta_r^2$ shown above in Figs.~6 and 7.
This is a consequence of the fact that $E_k$ is related to the spread of
the quantum paths, as measured by the mean-square displacements $Q_r^2$, 
and not to the centroid motion, which represents displacements of the
overall paths, irrespective of their size and shape.
The fast increase in $\Delta_r^2$ close to the spinodal points is mainly
due to an increase in the centroid delocalization $C_r^2$.

\section{Summary}

We have presented results of PIMD simulations of ice Ih in the
isothermal-isobaric ensemble at different pressures and temperatures.
This kind of simulations have allowed us to study this water phase in the
pressure region where it is metastable, and approach the spinodal points at
which it becomes mechanically unstable. At $P < 0$  we found
a spinodal pressure $P_s$ ranging from --1.4 to --0.7 GPa in the temperature
range from 50 to 300 K. At positive pressures, we obtained a spinodal
in the range $P_s' \sim$ 1.1--1.3 GPa, that is well defined at temperatures 
$T \lesssim$ 100 K.
At higher temperatures, kinetic processes favor amorphization of the solid 
for pressures clearly lower than the estimated spinodal pressure.

Although the molar volume decreases with increasing pressure, the interatomic
distances in the solid change in a peculiar way that remembers their 
temperature dependence.  Thus, the distance between oxygen atoms 
in neighboring molecules decreases for increasing pressure, but the 
intramolecular O--H distance becomes larger for higher pressure. 
This is a consequence of the hydrogen bonds connecting
contiguous molecules, which become stronger as the volume (or O--O distance)
is reduced, causing a weakening of the intramolecular O--H bonds. 

For the bulk modulus of ice Ih, we obtain a maximum at 0.3--0.5 GPa,
which increases slowly as the temperature is lowered. At higher and
lower (negative) pressures, $B$ is found to decrease and extrapolates
to zero at the corresponding spinodal pressure ($P_s$ or $P_s'$).  
Close to the spinodals, we observe an increase in the atomic
delocalization for both oxygen and hydrogen atoms, especially at
temperatures in the order of the melting temperature of ice Ih.
For the kinetic energy, however, we do not observe any anomalous
effect, and it is found to increase smoothly as temperature
or pressure is raised.

We have assessed the importance of quantum effects by comparing results
obtained from PIMD simulations with those yielded by classical
simulations. Concerning the stability of ice Ih, we have found that 
quantum effects reduce the metastability region of this solid phase,
for both positive and negative pressures. At low temperatures, this 
means that both spinodal pressures $P_s$ and $P_s'$  are shifted by 
about 0.3 GPa. 
Structural variables also change when quantum nuclear motion is
considered. Thus, the crystal volume, interatomic distances, and atomic
delocalization suffer appreciable modifications in the range of
temperature and pressure considered here. Most important is the
increase in the mean-square displacement of hydrogen, which contributes
to weaken the intermolecular H bonds, and to strengthen the
intramolecular O--H bonds.

An extension of this work can consist in studying the high-density
amorphous phase obtained from ice Ih under pressure, as well as other
amorphous water phases, which may show peculiar properties as a function of
pressure and temperature. Path-integral molecular dynamics simulations with
interatomic potentials similar to that employed here can be also adequate 
to study these solid phases.

\begin{acknowledgments}
This work was supported by Ministerio de Ciencia e Innovaci\'on (Spain)
through Grant FIS2009-12721-C04-04 and by Comunidad Aut\'onoma de Madrid 
through Program MODELICO-CM/S2009ESP-1691.
\end{acknowledgments}

\end{document}